\documentclass[fleqn,usenatbib]{mnras}

\usepackage{hyperref}
\usepackage{bm}
\usepackage{acro}
\usepackage[T1]{fontenc}
\DeclareRobustCommand{\VAN}[3]{#2}
\let\VANthebibliography\thebibliography
\def\thebibliography{\DeclareRobustCommand{\VAN}[3]{##3}\VANthebibliography}
\usepackage{graphicx}
\usepackage{amsmath}
\usepackage{amssymb}

\DeclareAcronym{LHAASO}{
    short = LHAASO,
    long = the Large High Altitude Air Shower observatory
}

\DeclareAcronym{HAWC}{
    short = HAWC,
    long = the High-Altitude Water Cherenkov Gamma-Ray Observatory
}

\title[Pulsars and $\gamma$-ray spectra of LHAASO]{Pulsars as candidates of LHAASO sources J2226+6057, J1908+0621 and J1825-1326}

\author[Z. Chang et al.]{
Zhe Chang,$^{1,2}$
Xukun Zhang,$^{1,2}$\thanks{E-mail: zhangxukun@ihep.ac.cn}
Jing-Zhi Zhou$^{1,2}$
\\
$^{1}$Institute of High Energy Physics, Chinese Academy of Sciences, Beijing 100049, China\\
$^{2}$University of Chinese Academy of Sciences, Beijing 100049, China
}

\date{Accepted XXX. Received YYY; in original form ZZZ}
\pubyear{2022}

\begin{document}
\label{firstpage}
\pagerange{\pageref{firstpage}--\pageref{lastpage}}
\maketitle

\begin{abstract}
    The LHAASO Collaboration has observed ultrahigh-energy photons up to $1.4$PeV from $12$ $\gamma$-ray Galactic sources. In particular, the $\gamma$-ray spectra of the sources J2226+6057, J1908+0621, J1825-1326 have been published. We investigate the possibility of suggested origin pulsars near the sources as the PeVatrons. The pulsar is described by a rotating magnetic dipole. Assuming protons are uniform distributed out of the light cylinders, we obtain the Lorentz distribution of proton energy spectrum.  It is found that the protons around pulsar could be accelerated to PeV at short times. The hadronic $\gamma$-ray spectra of the suggested origin pulsars are in good agreement with the LHAASO observed $\gamma$-ray spectra of the sources J2226+6057, J1908+0621, J1825-1326.
\end{abstract}


\section{Introduction}\label{sec:intro}
\iffalse

\begin{figure*}
    \centering
    \includegraphics[scale=0.15]{pulsar.png}
    \caption{The protons can be accelerated rapidly out of the light cylinder. There is a strong electromagnetic field which can accelerate protons to PeV at short times. The thick black line represents the rotation axis of the pulsar. The brown line is the magnetic axis. The blue cylinder denotes the light cylinder. The thin black curve is the electric field line. The orange and green curves correspond to the trajectories of protons.}\label{fig:pulsar}
\end{figure*}
\fi

The ultrahigh-energy photon is one of the most powerful probe for very-high-energy astrophysics. It contains a rich vein of information of astrophysics and fundamental physics. Recently, twelve sources in the Galaxy of the ultrahigh-energy photons have been observed by \ac{LHAASO} (\cite{cao_ultrahigh-energy_2021}). The energies of the observed photons range from $0.1\sim 1.4$PeV. Earlier, the Tibet AS$\gamma$ (\cite{Amenomori:2019rjd}), \ac{HAWC} (\cite{Abeysekara:2021yum, HAWC:2019tcx}) also have reported the detection of the protons around $0.1$PeV from the Galaxy. All of them indicate that there may exist unknown PeVatrons in the Galaxy. In particular, the $\gamma$-ray spectra of the sources J2226+6057, J1908+0621, J1825-1326 have been published by \ac{LHAASO}. The \ac{LHAASO} suggested possible origins are SNR G106.3+2.7 and PSR J2229+6114 for source J2226+6057; SNR G40.5-0.5, PSR J1907+0602, and PSR J1907+0631 for source J1908+0621; PSR J1826-1334 and PSR J1826-1256 for source J1826-1326.

The acceleration mechanisms of cosmic rays has been studied for a long time. One of the most popular acceleration mechanism is the diffusive shock acceleration related to the supernova remnants (\cite{Fermi:1949ee, Drury:1983zz, Schure:2012du}). Charged particles would be accelerated by the shock waves produced by the supernova explosion. The other possible origin of the high energy cosmic rays is the pulsar. For the pulsars, most acceleration mechanisms of cosmic rays are based on the properties of the magnetosphere structure and the radiation location, such as polar cap (\cite{Daugherty:1995zy}), slot gap (\cite{Dyks:2003rz, Muslimov:2004ig}) and outer gap (\cite{Cheng:1986qt}), which induce different spectra. The pulsar magnetosphere model is normally studied in terms of numerical simulations (\cite{Spitkovsky:2006np, Kalapotharakos:2008zc, Contopoulos:2009vm, Tchekhovskoy:2012hm, Philippov:2014mqa}).  

In this paper, we propose a concise and effective model to explain the LHAASO's observations on J2226+6057, J1908+0621, and J1825-1326, respectively. We obtain the energy distribution of the pulsar accelerated protons in terms of the acceleration mechanisms in \cite{Chang:2021bpc}. Then we use the energy distribution of the protons to calculate the $\gamma$-ray spectra from PSR J2229+6114, PSR J1907+0602, PSR J1907+0602, PSR J1907+0631, PSR J1826-1334 and PSR J1826-1256 by using the \texttt{naima} code (\cite{naima}). More precisely, in this model, the pulsar is described as a rotational magnetic dipole. The corresponding acceleration mechanism is based on the Hertzian magnetic dipole field of pulsar instead of the magnetosphere structure. In order to avoid the influence of the magnetosphere, we assume that the particles are initially located out of the light cylinder. They are accelerated by the rotating magnetic dipole field of the pulsar. We find that the protons around pulsar out of the light cylinder could be accelerated to PeV at short times and the energy distribution of the pulsar accelerated protons is Lorentz distribution with adjusted $R^2$ larger than $0.9$. We use this Lorentz distribution of protons to calculate the $\gamma$-ray spectra. It is found that the $\gamma$-ray spectra of PSR J2229+6114, PSR J1907+0602, PSR J1907+0631, RSR J1826-1334, PSR J1826-1256 fit well with \ac{LHAASO}'s observations on J2226+0657, J1908+0621, and J1825+1326.

The remaining part of this paper is organized as follows. In Sec.~\ref{sec:cal}, we describe the pulsar by a Hertzian magnetic dipole, and show that a proton around the pulsar out of light cylinder could be accelerated to a high energy. In Sec.~\ref{sec:energy_spectrum}, the acceleration of uniform distributed protons around the pulsars are simulated. We obtain the $\gamma$-ray spectra of pulsars through the hadronic process. Our main conclusions are finally summarized in Sec.~\ref{sec:con}.

\section{Particle accelerated by a Hertzian magnetic dipole}\label{sec:cal}

In this section, we briefly review the key features of the Hertzian magnetic dipole field of pulsar. We study the motion of protons in the electromagnetic field of pulsar. It shows that the protons around pulsar out of the light cylinder could be accelerated to PeV at short times. As shown in Ref.~\cite{Chang:2021bpc}, pulsar can be regarded as a point-like rotational magnetic dipole $\bm{M}$ out of light cylinder, which can be described by the solution of the d'Alembert equation,
\begin{equation}\label{eq:A}
	\begin{gathered}
		\phi(t, \bm{x})=0 \ , \\
		\bm{A}(t, \bm{x})=\frac{\mu_0}{4\pi}\left(\bm{M}+\frac{r}{c} \dot{\bm{M}}\right)_{\mathrm{ret}} \times \frac{\bm{r}}{r^{3}} \ ,
	\end{gathered}
\end{equation}
where $c$ is the speed of light, $\mu_0$ denotes the permeability of vacuum, $\dot{\bm{M}}=\bm{\Omega} \times \bm{M}$, and $\bm{\Omega}$ is the rotational velocity of pulsars. As shown in Eq.~(\ref{eq:A}), the retarded potential $\bm{A}$ can be evaluated by rotational magnetic dipole $\bm{M}_{\mathrm{ret}} \equiv \bm{M}(t-r /c)$. The electromagnetic fields can be calculated as
\begin{equation}\label{eq:B}
	\begin{aligned}
		\bm{B}(t, \bm{x})=& \nabla \times \bm{A}=\frac{\mu_0}{4\pi}\left(-\frac{\bm{M}}{r^{3}}-\frac{1}{r^{2} c} \dot{\bm{M}}-\frac{1}{r c^{2}} \ddot{\bm{M}}+\frac{3 \bm{r}}{r^{5}}(\bm{r} \cdot \bm{M})\right.\\
		&\left.+\frac{3 \bm{r}}{r^{4}}\left(\bm{r} \cdot \frac{1}{c} \dot{\bm{M}}\right)+\frac{\bm{r}}{r^{3}}\left(\bm{r} \cdot \frac{1}{c^{2}} \ddot{\bm{M}}\right)\right)_{\mathrm{ret}} \ ,
	\end{aligned}
\end{equation}

\begin{equation}\label{eq:E}
	\bm{E}(t, \bm{x})=-\nabla \phi-\frac{\partial \bm{A}}{\partial t}=-\frac{\mu_0}{4\pi}\left(\dot{\bm{M}}+\frac{r}{c} \ddot{\bm{M}}\right)_{\mathrm{ret}} \times \frac{\bm{r}}{r^{3}} \ .
\end{equation}
In the electromagnetic field, a charged particle with charge $q$ and mass $m$, obeys the Landau-Lifshit (LL) equation (\cite{book:14839, Price:2021zqq}),
\begin{equation}\label{eq:eq_of_motion}
    ma^\mu = eF^{\mu\nu}u_\nu + \tau_0 \left( q\frac{{\rm d}F^{\mu\nu}}{{\rm d}\tau}u_\nu + \frac{q^2}{m}P^\mu_\nu F^{\nu\alpha}F_{\alpha\beta}u^\beta\right) \ ,
\end{equation}
where
\begin{eqnarray}
    P^\mu_\nu = \delta^\mu_\nu - \frac{u^\mu u_\nu}{c^2} \ , \\
    \tau_0 = \frac{2}{3}\frac{q^2}{4\pi\epsilon_0 m c^3} \ ,
\end{eqnarray}
$F_{\mu\nu}$ is the electromagnetic tensor, $u_{\mu}$ is $4$-velocity of the particle, and $\epsilon_0$ represents the vacuum dielectric constant. The $\tau_0$-terms denote the radiation reaction force, which describe the influence of the particle's own electromagnetic field. Using Eq.~(\ref{eq:B})-(\ref{eq:eq_of_motion}) we obtain the accelerating information for a charged particle near a pulsar out of light cylinder. 

\begin{figure}
    \centering
    \includegraphics[scale=0.45]{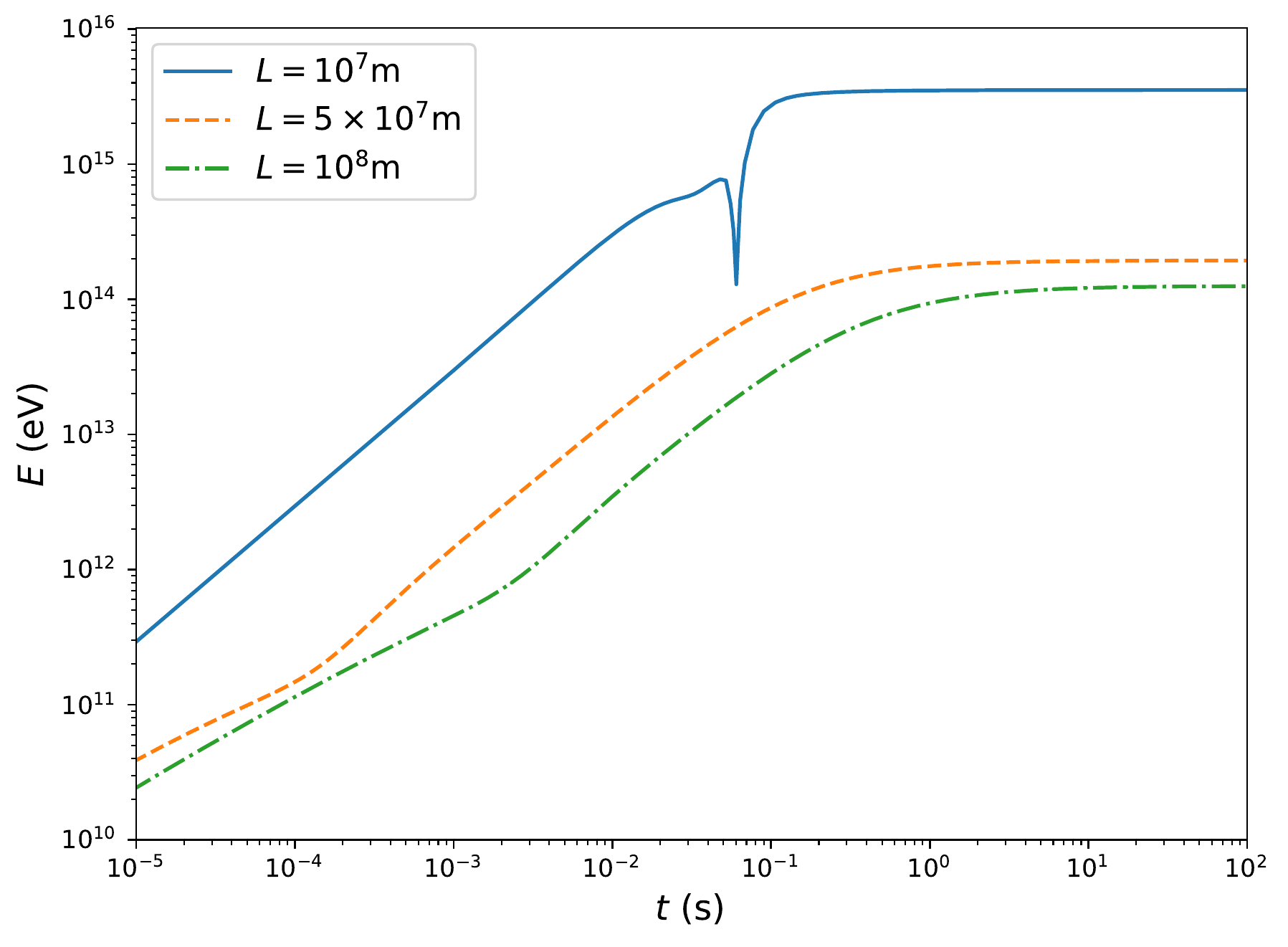}
    \caption{The accelerating process of protons. The horizontal axis represents the coordinate time $t$, while the vertical one denotes the kinetic energies in units of eV. Here, we have set the inclination angle of pulsars $\theta = \pi/6$, angular speed $\Omega = 20\pi\ \mathrm{s}^{-1}$ and magnetic moment $M = 4\times 10^{27} \mathrm{Am}^2$. The three protons are located at the equatorial plane with distance $10^7$m (blue solid curve), $5\times10^7$m (orange dashed curve) and $10^8$m (green dot-dashed curve), respectively.}\label{fig:accelerate_curve}
\end{figure}

Fig.~\ref{fig:accelerate_curve} shows the accelerating processes of three protons located initially at the equatorial plane of the pulsar with different distances. All protons are out of the light cylinder. We obtain that the Hertzian magnetic dipole field of the pulsar could accelerate the protons to high energies in a short time. For the pulsar with high angular speed and strong magnetic moment, the proton located closer to the pulsar could be accelerated to PeV energy (blue curve).

\section{The hadronic $\gamma$-ray spectrum of pulsar}\label{sec:energy_spectrum}
% 1. 假设, 质子谱图,   2. 质子碰撞, 光子谱图

\begin{table}
    \centering
    \caption{Some parameters of the five pulsars. Here, $P$ denotes the period of the pulsar, $\dot{P}$ is its first derivative, $B_s$ represents the surface magnetic field and $\theta$ is the inclination angle. These data come from \protect\cite{Halpern:2001fc, Abdo:2010ht, Lyne:2016kam, Duvidovich:2019ykz, Fermi-LAT:2013svs}, respectively.}\label{tab:inf}
    \begin{tabular}{lccr}
		\hline
		PSR & $P$ (ms) & $\dot{P}\ (\mathrm{s}\cdot\mathrm{s}^{-1})$  & $B_s \sin\theta$ (G)\\
		\hline
		J2229+6114& $51.3$ & $7.83\times 10^{-14}$ & $2.1\times 10^{12}$\\
		J1907+0602& $107$ & $8.68\times10^{-14}$ & $3.0\times10^{12}$\\
		J1907+0631& $324$ & $4.52\times10^{-13}$ & $1.2\times10^{13}$\\
        J1826-1334& $101$ & $7.52\times 10^{-14}$ & $2.8\times 10^{12}$\\
        J1826-1256& $110$ & $1.21 \times 10^{-13}$ & $3.7 \times 10^{12}$\\
		\hline
	\end{tabular}
\end{table}

Ref.~\cite{cao_ultrahigh-energy_2021} gives $\gamma$-ray spectra of the three sources LHAASO J2226+6057, J1908+0621 and J1825-1326. All of them have potential pulsar sources, PSR J2229+6114 for J2226+6057, PSR 1907+0602, PSR J1907+0631 for J1908+0621 and PSR J1826-1334, PSR J1826-1256 for J1825-1326.  In this section, we use the acceleration mechanism in Sec.~\ref{sec:cal} to calculate the energy distribution of protons. It shows that the energy distribution of the pulsar accelerated protons is Lorentz distribution with adjusted $R^2$ larger than $0.9$. Then we use the energy distribution of the protons to calculate the $\gamma$-ray spectra from the above five pulsars by using the \texttt{naima} code (\cite{naima}).

\begin{figure*}
    \centering
    \includegraphics[scale=0.56]{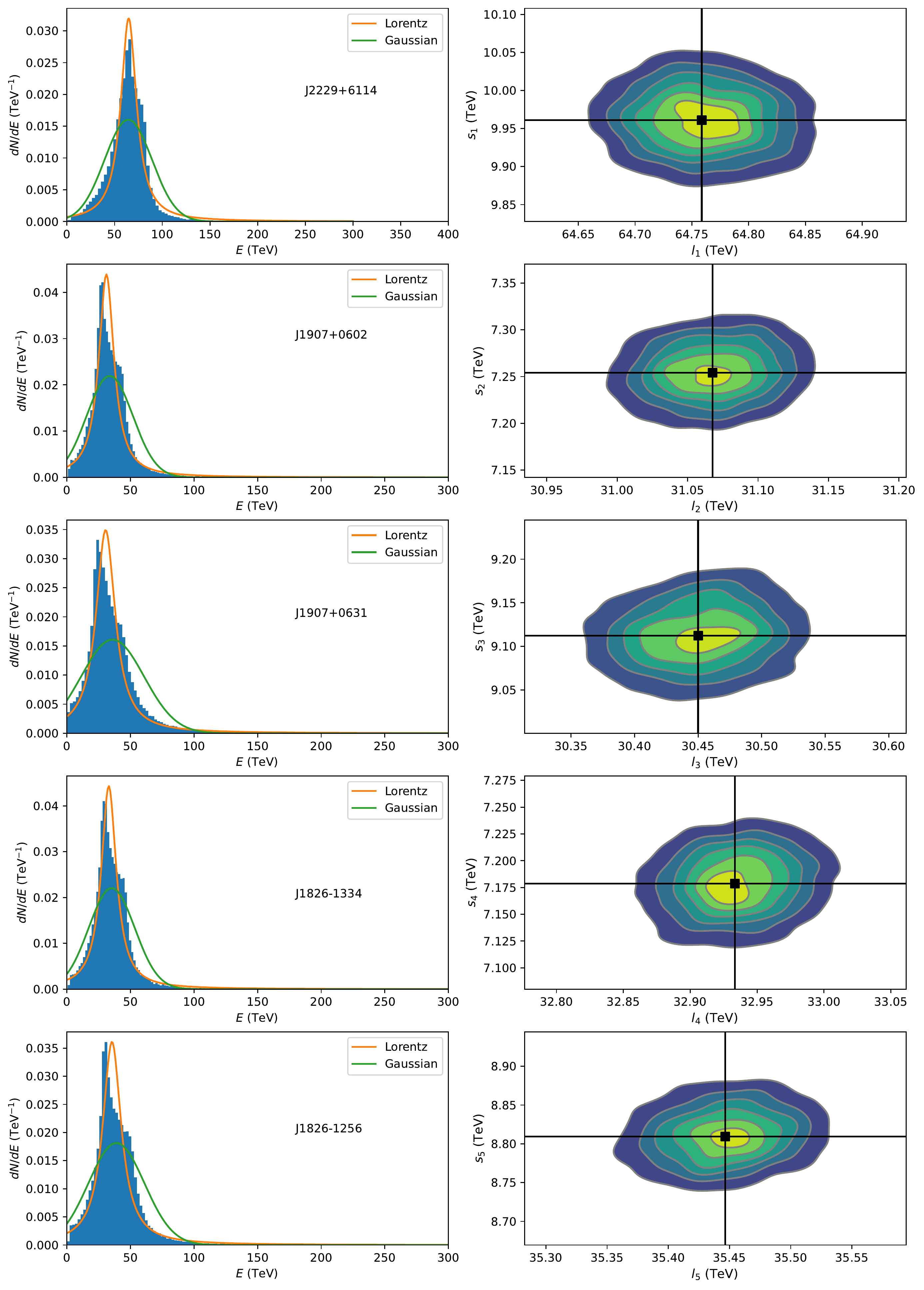}
    \caption{The energy distribution of the protons accelerated by five pulsars. For each pulsar adjacent to the $\gamma$-ray source, we give the histogram a Lorentz fit (orange curve) and a Gaussian fit (green curve) in the left panel. We find that the Lorentz models fit the histograms better and the adjusted $R^2$ are larger than 0.9. In right panel we give the posterior distributions for the parameters $l$ and $s$ of the Lorentz model.}\label{fig:proton_spectrum}
\end{figure*}

We assume that the protons are uniformly distributed in a spherical shell between $10^7 \sim 10^9$m around the pulsar. For a typical pulsar with mass $\mathcal{M} = 1.4M_\odot$, period $P = 0.1$s and magnetic dipole $|\bm{M}| = 4\times10^{27} \mathrm{Am}^2$, protons in the spherical shell are dominated by the electromagnetic force and safely outside the light cylinder. We set the radius of the pulsar $10$ km and the inclination angle $\theta = \pi/6$. Then the protons are accelerated by the electromagnetic field of the pulsars. Using the parameters in Tab.~\ref{tab:inf}, we obtain the simulated results in Fig.~\ref{fig:proton_spectrum}. Here, the spectra of protons are fitted by the Lorentz distribution with adjusted $R^2$ larger than 0.9, i.e.,
\begin{equation}\label{lorentz_dis}
    \frac{dN_p}{dE_p} \propto \frac{1}{\pi s}\frac{1}{1+\left((E_p - l)/s\right)^2} \ ,
\end{equation}
where $s$ and $l$ are the free parameters characterize the location and scale of the distribution. 

From Fig.~\ref{fig:proton_spectrum} we obtain that the protons around the pulsars would be accelerated to high energies and obey the Lorentz distribution. Such ultrahigh-energy protons would collide with hydrogen in ambient medium and produce secondary $\pi$ and $\eta$ mesons. Subsequently, photons are produced by the decay reaction $\eta \rightarrow 2\gamma$, $\eta \rightarrow \pi^+ \pi^- \pi^0$, $\eta \rightarrow 3\pi^0$ and $\pi^0\rightarrow \gamma\gamma$.
%Subsequently, these mesons decay to high-energy leptons. 
The process of inelastic proton-proton (p-p) collisions has been studied (\cite{Kelner:2006tc, Kafexhiu:2014cua}). The spectrum of the gammy ray is given by
\begin{equation}
    \frac{{\rm d}F}{{\rm d}E_\gamma} = \frac{c n_H}{4\pi l^2}\int_{E_\gamma}^{E_\mathrm{max}} \sigma_\mathrm{inel}(E_p)\frac{{\rm d}N_p}{{\rm d}E_p}F_\gamma \left(\frac{E_\gamma}{E_p}, E_p\right) \frac{{\rm d}E_p}{E_p} \ ,
\end{equation}
where $E_\mathrm{max}$ is the maximum energy of the protons accelerated by the pulsar, $l$ is the distance from the pulsar to the earth, $n$ is the density of the target particle, $\sigma_\mathrm{inel}$ is the cross section of the inelastic p-p interactions, and the function $F_\gamma$ is defined in Ref.~\cite{Kelner:2006tc}. We calculate the spectrum using the \texttt{naima} code (\cite{naima}). 

By acceleration and collisions, some protons are consumed. New protons could be continuously replenished by the plasma wind of the magnetosphere of the pulsar (\cite{Goldreich:1969sb, 1971ApJ...164..529S, 1975ApJ...196...51R, philippov_pulsar_2022}). The dynamic equilibrium of the protons are built in terms of the hadronic pulsar wind from the magnetosphere and the interactions of protons out of the light cylinder.

\begin{figure*}
    \centering
    \includegraphics[scale=0.5]{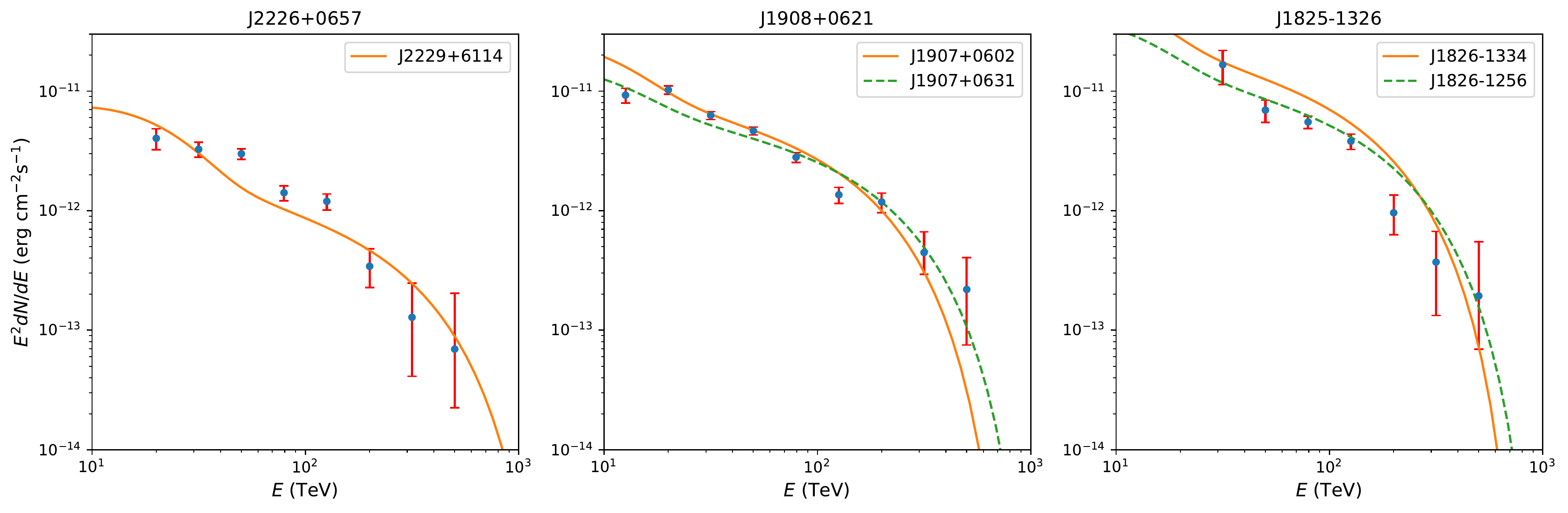}
    \caption{The $\gamma$-ray spectra of our model. Blue points are the observations of the \ac{LHAASO} (\protect\cite{cao_ultrahigh-energy_2021}). The orange solid curves and the green dashed curves denote the spectra of the inelastic p-p collisions around the pulsars near the radiation sources. The PSR J2229+6114 for source J2226+6057; PSR J1907+0602, and PSR J1907+0631 for source J1908+0621; PSR J1826-1334 and PSR J1826-1256 for source J1826-1326.}\label{fig:result}
\end{figure*}

The $\gamma$-ray spectra from PSR J2229+6114, PSR J1907+0602, PSR J1907+0602, PSR J1907+0631, PSR J1826-1334, PSR J1826-1256 are shown in Fig.~\ref{fig:result}. We find that the observed $\gamma$-ray spectra of LHAASO can be explained by our model as the hadronic spectra of pulsars. The protons out of the light cylinder of the pulsars are accelerated by a Hertzian magnetic dipole. Then the protons collides with the hydrogen in ambient medium and high-energy $\gamma$-rays are produced.

\section{Conclusion and Discussion}\label{sec:con}
% 1. 做的事,  2. 磁层忽略  2.1 磁层影响   3. 未来

In this paper, we used a Hertzian magnetic dipole to describe the electromagnetic field around the pulsars. For the five pulsars near the sources LHAASO J2226+6057, J1908+0621, J1825-1326, we found that the protons around the pulsars could be accelerated to PeV energies. The energy distribution of protons after accelerated obey a Lorentz distribution. 
%The Lorentz distribution can be approximated by the power law spectrum with spectral index $n=-2$ when the energies approach to PeV. This is in agreement with observation on energy spectrum of cosmic rays (\cite{Blumer:2009jrd}). The spectral index is about $-2.7$ at energies up to several PeV, and steepening with $-3.1$ at higher energies. In our model, the largest energy of accelerated protons is about PeV, this may be the origin of the steep of the spectrum. 
Through hadronic processes, eventually, the $\gamma$-ray spectra in Fig.~\ref{fig:result} are obtained. It shows that our model could explain the observations of LHAASO well. Here, we only considered the hadronic spectra. For the leptonic $\gamma$-ray production, the energy loss should be taken account (\cite{cao_ultrahigh-energy_2021}), which may be studied in the future.

\begin{figure*}
    \centering
    \includegraphics[scale=0.5]{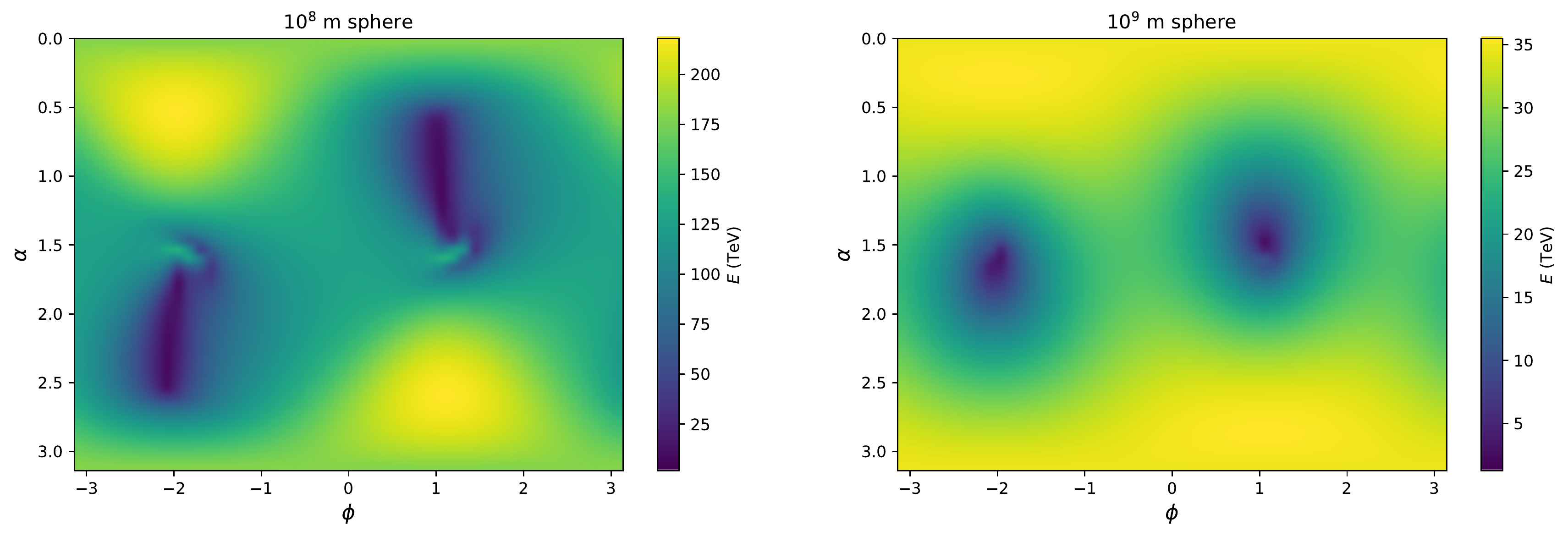}
    \caption{The final energy of protons related to the initial locations on the $10^8$m (left panel) and $10^9$m (right panel) sphere surfaces of the pulsar. The rotation axis is chosen as $z$ axis. $\alpha \in [0, \pi]$ is the polar angle. $\phi \in [-\pi, \pi]$ is the azimuthal angle. We have set the inclination angle of the pulsar $\theta=\pi/6$, angular speed $\Omega=20\pi$ and magnetic dipole $M=4\times10^{27}\mathrm{Am}^2$.}\label{fig:1e8sphere}
\end{figure*}

%Fig.~\ref{fig:1e8sphere} gives the final energy of protons related to the initial locations on the $10^8$m and $10^9$m sphere surfaces of a pulsar. In principle, one can obtain any aimed proton spectrum (and therefore the $\gamma$-ray spectrum) by modifying the distribution of the initial protons. However, the distribution must be reasonable.  Note for the protons initially located on the $10^9$m sphere, the largest final energy accelerated is about $35$ TeV and contribute little to our results. Thus, in this paper, we only consider the protons distributed in a spherical shell between $10^7 \sim 10^9$m around the pulsar. In this region the protons are accelerated to high energies and out of the light cylinder. 
%At the same time, due to the fact that protons are dominated by the electromagnetic force instead of gravity, we choose the uniform distribution.
Fig.~\ref{fig:1e8sphere} gives the final energy of protons related to the initial locations on the $r = 10^8$m and $10^9$m sphere surfaces of a pulsar. Note for the protons initially located on the $10^9$m sphere, the largest final energy accelerated is about $35$ TeV and contribute little to our results. Thus, in our model, we only consider the protons distributed in a spherical shell between $10^7 \sim 10^9$m around the pulsar. In this region the protons are accelerated to high energies and out of the light cylinder.

\begin{figure*}
    \centering
    \includegraphics[scale=0.45]{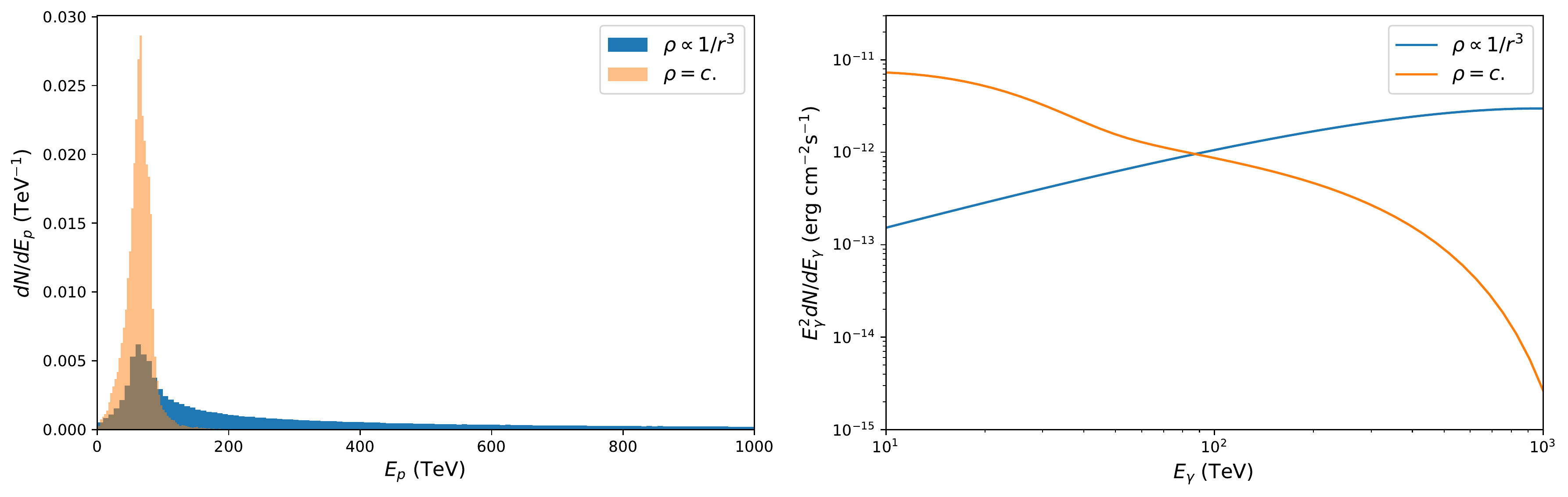}
    \caption{The final energy distribution of protons (left panel) and corresponding $\gamma$-ray spectra (right panel). The results
    of two initial spatial distributions are shown, $\rho \propto 1/r^3$ (blue) and $\rho =$ constant (orange). All protons are initially
    distributed in the spherical shell between $10^7 \sim 10^9$ m around PSR J2229+6114.}\label{fig:dis_compare}
\end{figure*}

The relation between different initial proton spatial
distributions and the $\gamma$-ray spectra is considered. If the initial protons distribute denser in the yellow ($\alpha, \phi$-dim) or small-$r$ region in Fig.~\ref{fig:1e8sphere}, the ratio of high energy protons will increase and make the spectra of protons and resulted $\gamma$-ray harder. As an example, in Fig.~\ref{fig:dis_compare}, two spectra of protons and resulted $\gamma$-ray from different initial proton spatial distributions are shown. In the left panel, the blue histogram denotes the final energy spectrum of protons ${\rm d}N/{\rm d}E_p$ with an initial spatial distribution $\rho \propto 1/r^3$. 
%where $r$ is the distance between the proton and the center of pulsar. 
The orange histogram represents ${\rm d}N/{\rm d}E_p$ with an initial uniform distribution ($\rho = $ constant). The right panel gives corresponding $\gamma$-ray spectra. Compared with the uniform distribution, the distribution of $\rho \propto 1/r^3$ makes more protons locate closer to the pulsar and achieve higher energies by acceleration. As a result, the spectra of protons and $\gamma$-ray become harder. 
On the contrary, one can make the proton and $\gamma$-ray spectra softer by make the initial protons distribute denser in the blue ($\alpha, \phi$-dim) or large-$r$ region in Fig.~\ref{fig:1e8sphere}. In this paper, for simplicity, we assume a uniform initial distribution of protons.
%due to the fact that the protons in the shell are dominated by the electromagnetic force instead of gravity, we choose the uniform distribution.}

%A possible magnetosphere structure around the pulsar is neglected in our model. 
To minimize the impact of the magnetosphere structure, we have set the initial positions of the protons safely out of the light cylinder, i.e., about $10^7$m for the five pulsars. For a proton located at a smaller radius outside the light cylinder, it may be accelerated to a higher energy. However, the magnetosphere structure will inevitably affect the outside electromagnetic field. The acceleration of the protons nearby the light cylinder should be studied in the future.

\section*{Acknowledgement}
We thank Dr. Q.H. Zhu, Prof. Y. Chen, and Prof. Q. Zhao for useful discussions. This work has been funded by the National Nature Science Foundation of China under grant No. 12075249 and 11690022, and the Key Research Program of the Chinese Academy of Sciences under Grant No. XDPB15.

\section*{Data availability}

The LHAASO data is available at \href{http://english.ihep.cas.cn/lhaaso/pdl/202110/t20211026_286779.html}{http://english.ihep.cas.cn/lhaaso/pdl/202110/t20211026\_286779.html}. The parameters of the five pulsars studied in this paper can be found in \cite{Halpern:2001fc, Abdo:2010ht, Lyne:2016kam, Duvidovich:2019ykz, Fermi-LAT:2013svs}.

\bibliographystyle{mnras}
\bibliography{biblio}

\bsp
\label{lastpage}
\end{document}